\documentclass[12pt,preprint]{aastex}
\begin{document}

\title{Further Results from the Galactic O-Star Spectroscopic Survey:
Rapidly Rotating Late ON Giants}

\author{Nolan R.\ Walborn}
\affil{Space Telescope Science Institute,$^1$ 3700 San Martin Drive, Baltimore, MD 21218}
\email{walborn@stsci.edu}

\author{Jes\'us Ma\'{\i}z Apell\'aniz,$^2$ Alfredo Sota,\ Emilio J.\ Alfaro}
\affil{Instituto de Astrof\'{\i}sica de Andaluc\'{\i}a--CSIC, Glorieta de la
Astronom\'{\i}a s/n, 18008 Granada, Spain}
\email{jmaiz@iaa.es, sota@iaa.es, emilio@iaa.es}

\author{Nidia I.\ Morrell}
\affil{Las Campanas Observatory, Observatories of the Carnegie Institution 
of Washington, Casilla 601, La Serena, Chile}
\email{nmorrell@lco.cl}

\author{Rodolfo H.\ Barb\'a,$^3$ Julia I.\ Arias}
\affil{Departamento de F\'{\i}sica, Universidad de La Serena, Cisternas
1200 Norte, La Serena, Chile}
\email{rbarba@dfuls.cl, julia@dfuls.cl}

\author{Roberto C.\ Gamen}
\affil{Instituto de Astrof\'{\i}sica de La Plata--CONICET and Facultad de
Ciencias Astron\'omicas y Geof\'{\i}sicas, Universidad Nacional de La Plata, 
Paseo del Bosque s/n, 1900 La Plata, Argentina}
\email{rgamen@fcaglp.unlp.edu.ar}

\altaffiltext{1}{Operated by the Association of Universities for Research in Astronomy, Inc., under NASA contract NAS5-26555.}
\altaffiltext{2}{Visiting Astronomer, Las Campanas Observatory.}
\altaffiltext{3}{Also Instituto de Ciencias Astron\'omicas 
de la Tierra y del Espacio (ICATE--CONICET), Avenida Espa\~na 1512 Sur,
J5402DSP, San Juan, Argentina.  Visiting Astronomer, LCO.}

\begin{abstract}
With new data from the Galactic O-Star Spectroscopic Survey, we confirm and 
expand the ONn category of late-O, nitrogen-enriched (N), rapidly rotating 
(n) giants.  In particular, we have discovered two ``clones'' (HD~102415 and 
HD~117490) of one of the most rapidly rotating O stars previously known 
(HD~191423, ``Howarth's Star'').  We compare the locations of these objects in 
the theoretical HR Diagram to those of slowly rotating ON dwarfs and 
supergiants.  All ON giants known to date are rapid rotators, whereas no ON 
dwarf or supergiant is; but all ON stars are small fractions of their 
respective spectral-type/luminosity-class/rotational subcategories.  The ONn 
giants, displaying both substantial processed material and high rotation at an 
intermediate evolutionary stage, may provide significant information about 
the development of those properties.  They may have preserved high initial 
rotational velocities or been spun up by TAMS core contraction; but
alternatively and perhaps more likely, they may be products of binary mass 
transfer.  At least some of them are also runaway stars.

\end{abstract}


\section{Introduction}

Massive stars burn hydrogen to helium on the CNO cycle(s); the slowest
reactions have the effect of leaving nearly all of these elements as
nitrogen, at the outset and for their duration.  If this partially
(since the conversion of H to He proceeds on the nuclear timescale)
processed material is mixed to the surface before the reactions run
to completion, enhanced nitrogen to carbon and oxygen abundance ratios 
become visible.  Thus, direct observations of products from nuclear 
reactions inside the stars can be made and may provide vital evolutionary 
diagnostics, if they can be correctly interpreted.  

Early on it was realized that Wolf-Rayet stars of the nitrogen sequence
(WN) are revealing the products of H-burning, while the carbon sequence
(WC) is more evolved and shows He-burning products (see Crowther 2007 for
a recent review and references).  However, the phenomenology (let alone
the physics) is highly complex: for instance, very massive stars may
become WN solely through internal mixing and mass loss; at intermediate 
masses, WN stars may be post-red supergiants; and mass transfer
in binary systems can produce WN objects at all masses down to at least
20~$M_{\sun}$.  More recently, it has been shown that the timescale and
degree of these effects are functions of the initial stellar rotational 
velocity (Maeder \& Meynet 2000; Heger \& Langer 2000). 

An analogous dichotomy among certain absorption-line OB spectra was described
by Walborn (1971a), who adopted a classification notation analogous to
that for WR spectra, i.e. OBN and OBC.  However, it soon became apparent
from their strong hydrogen lines and space distribution that the latter
category is {\it less\/} evolved than the former, which in turn led to the
suggestion that the morphologically normal majority of OB supergiants may
already display some mixing of processed material in their spectra, while
the OBN are extreme cases (Walborn 1976).  This interpretation has
received substantial subsequent support (Smith \& Howarth 1994; Maeder \&
Conti 1994).

The OBN/OBC phenomenology is also complex and likely has a range of origins 
among different objects, just as in the WR case.  Moreover, its detectability 
is a strong function of the available spectroscopic criteria, which vary as 
a function of spectral type; the late-O and early-B range is particularly 
favorable, because of the number of comparably strong C, N, and O features 
present, which are strongest in the supergiants.  Examples in high-quality 
digital data can be found in Walborn \& Fitzpatrick (1990), Walborn \& 
Howarth (2000), and Walborn (2009a).  Quantitative measurements and 
analysis can mitigate these selection effects to a considerable extent.  
Recent investigations from different perspectives have been presented by 
Fraser et~al.\ (2010), Przybilla et~al.\ (2010), and Lyubimkov et~al.\ (2010).  
Rotation is again a key parameter, as is the possibility of angular momentum 
and abundance alterations by mass transfer in binary systems (Langer et~al.\  
2008).  These effects may produce systematic differences in mixing among 
some clusters and associations.

Howarth \& Smith (2001) showed that the remarkable late-O giant HD~191423 
(``Howarth's Star'') is one of the two most rapidly rotating massive stars 
analyzed so far and displays a strong nitrogen enhancement in its spectrum; 
thus, it combines two of the key parameters of interest in the present context.
Walborn (2003) included it in a small, newly defined category of ONn spectra, 
where the ``N'' denotes nitrogen enhancement while ``n'' (for ``nebulous'') is 
the classical descriptor of broadened absorption lines.  Here we report the 
discovery of two extremely similar spectra and improve the spectral types of
several other members of the ONn category, thus expanding its membership
and confirming its existence.  We also further discuss the characteristics 
and possible significance of this category.

\section{Observations}

The Galactic O-Star Spectroscopic Survey (GOSSS, Ma\'{\i}z Apell\'aniz et~al.\  2010), including its data and analysis procedures, is extensively 
discussed by Sota et~al.\ (2011), which will not be repeated in detail here.  
Briefly, GOSSS is a systematic, uniform survey of the blue-violet spectra 
of all accessible Galactic O stars, with high-S/N (200--300) digital data
covering at least 3900--5000~\AA\ at moderate resolution (R~$\sim2500$).  
As discussed in more detail below, this work has improved the definition of 
the spectral-classification system and it is revealing numerous objects and 
categories of special interest (e.g., Walborn et~al.\ 2010).  The present 
results constitute another example of the latter.  Several objects discussed 
here pertain to future GOSSS installments covering the southern hemisphere.

\section{Results}

A key spectral segment in the full current membership of the ONn class is
presented in Figure~1, with the prototype subgiant ON spectrum of 
HD~201345 for comparison (Walborn 1970; its spectral type has been refined 
from the original in the GOSSS data). The defining ON morphology of N~III
$\lambda\lambda$4634, 4640-4642 and the C~III $\lambda$4650 blend, with 
the N~III $\lambda\lambda$4640-4642 blend dominant, is clearly seen.  As
shown by the further comparison object HD~96264, an O9.5~III standard, in
morphologically normal spectra the C~III feature is far stronger than the
N~III.  (In OC spectra it is even more so.)

\begin{figure}
\epsscale{0.63}
\plotone{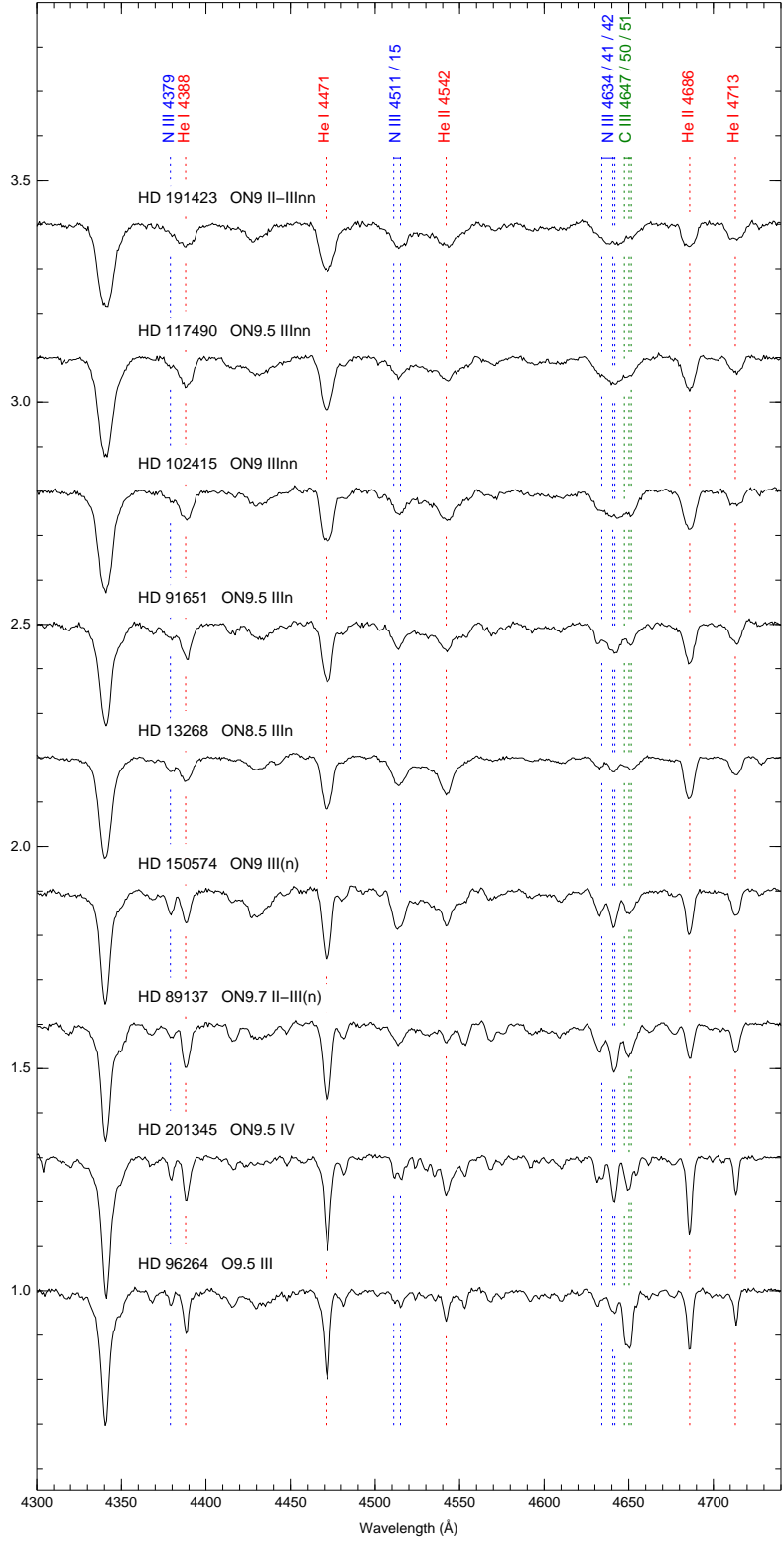}
\caption{
A $v\sin i$ sequence of ONn giant spectra in the blue-green region, with 
the ON subgiant HD~201345 and the normal standard HD~96264 for comparison.  
The ordinate is rectified continuum intensity.}
\end{figure}

It can also be readily seen that the absorption lines of all the classes
III and II objects are appreciably broadened to various degrees, and the
progressive effects on the N~III/C~III blends are clear.  Howarth
\& Smith (2001) showed how the spectrum of HD~191423, including the CNO
anomalies, is perfectly matched by a rotationally broadened version of
that of HD~201345.  They derived $v\sin i$~=~435~km~s$^{-1}$ for
HD~191423, the largest value found for an O star.\footnote{Herrero et~al.\  (1992) derived 450~km~s$^{-1}$ for this same star; see also Villamariz
et~al.\ (2002) further discussed below.}  In the rotational
line-broadening measures calculated by the digital classifier developed
by J.M.A.\ for GOSSS, the descriptors introduced by Walborn (1971b) have
the following correspondences: (n) to 200, n to 300, and nn to
400~km~s$^{-1}$, in excellent agreement with the directly determined
value for HD~191423, as well as those for several other objects in the 
figure listed by Walborn (2003).  Figure~1 shows the remarkable similarity
between the previously undiscussed spectra of the southern stars
HD~102415 and HD~117490, and that of HD~191423, including the extreme line
broadening. 

As discussed in more detail by Sota et~al.\ (2011), the quality and size
of the GOSSS dataset have motivated some refinements in the spectral
classification system and revisions to the spectral types of a number of 
individual stars.  These developments affect the ONn sample in the
following specific ways.  (1)~The definitions of types O8.5, O9, and O9.5   
have been standardized at all luminosity classes in terms of the line
ratios He~I~$\lambda$4144/He~II~$\lambda$4200 and 
He~I~$\lambda$4387/He~II~$\lambda$4541, which both have values of 
unity at O9 and deviate from that in opposite senses to either side.
(2)~The O9.7 spectral type, based upon a value near unity of the ratio
He~II~$\lambda$4541/Si~III~$\lambda$4552, is now used uniformly at all
luminosity classes; previously it had been defined only for classes I and II.  
That revision has been implemented by shifting some B0 and B0.2 types of 
classes III-V, including standards, to the next earlier type; thus the 
consistency of the horizontal types has been improved across the vertical 
classes.  (3)~The ON defining morphology in the adjacent N~III $\lambda$4640 
and C~III $\lambda$4650 blends was often difficult to discern with 
confidence in rapid rotators on photographic spectrograms because of the 
low contrast.  The high-S/N digital data improve this definition
substantially, allowing several previously suspected but uncertain cases
to be confirmed here; some historical notes for individual stars are given
with a table below.  (4)~The same uncertainty affected luminosity classes
based primarily upon He~II~$\lambda$4686/He~I~$\lambda$4713 in rapid
rotators; it is greatly alleviated in the digital data, which has led to   
some other revisions. 

The digital classifier used with the GOSSS data further contributes to
improving the systematic and random reliability of the results, because it 
compensates for the tendency of the eye to be drawn to central depths rather 
than equivalent widths in digital data.  That of course can lead to systematic 
effects as a function of line broadening or resolution.  The classifier 
displays any sequence of standard spectra consecutively, either overplotted 
or displaced from the unknown on the screen.  Moreover, the line widths (and 
consequently, depths) in the standards can be adjusted and matched to those of 
the unknown in terms of the n-parameter discussed above.  The program actually 
computes composite Doppler profiles for a rotating disk. Second-order effects 
such as limb darkening, oblateness, or non-rotational broadening are not 
included, as appropriate for the relatively low resolution of the GOSSS data.  
High-resolution spectroscopic data for as many of these stars as feasible are 
being obtained in the context of the associated OWN (Barb\'a et~al.\ 2010; 
Arias et~al.\ 2010) and IACOB (Sim\'on-D\'{\i}az et~al.\ 2011) programs and will 
receive full quantitative analyses.

Some parameters of the ONn stars, together with a substantial number of other 
ON, normal, and OC objects, are listed in Table~1.  All of the original ON/OC 
stars from Walborn (1976) are included here, together with a substantial 
number of morphologically normal comparison spectra of the same two-dimensional types, as an optimum sample for future quantitative analysis.  Note particularly three very rapidly rotating giants/subgiant {\it without\/}
nitrogen enhancement, all of which are also runaway stars (see Section
4.2 below)!  It will be of some interest to understand their evolutionary
histories in contrast to those of the ONn stars.  Many of these comparison 
spectral types have also been refined or newly derived in GOSSS, analogously
to the discussion of the ONn class above. The absolute visual 
magnitudes are from Walborn (1973), while the effective temperatures and 
bolometric corrections (albeit for nonrotating stars) are from Martins, 
Schaerer, \& Hillier (2005); values for spectral types O9.7--B0 and
luminosity class IV have been extrapolated or interpolated, respectively.  
Projected rotational velocities for the ON sample are given in Table~2,
along with radial velocities, galactic latitudes, nitrogen abundance
determinations as available, and notes on previous spectral classifications.
Two additional Galactic ON supergiants are excluded here: HD~105056, which 
may be a low-mass (PAGB?) object (Walborn, Conti, \& Vreux 1980); and 
BD~+36$^{\circ}$~4063, which is a short-period, spectrum-variable interacting 
binary (Howarth 2008; Williams et~al.\ 2009).

\begin{table}
\caption{Observational and Physical Parameters for ON, OB, and OBC Stars} 
\vspace{0.2cm}
\begin{tabular}{llrlcrlll}
\tableline
\tableline
\noalign{\smallskip}
\multicolumn{1}{c}{Name}&R.A.(2000.0)&Decl(2000.0)&Spectral
Type&$V$&$B-V$&\multicolumn{1}{c}{$M_V$}&\multicolumn{1}{c}{$T_{\rm
eff}$}&\multicolumn{1}{c}{BC}\\
\noalign{\smallskip}
\tableline
\noalign{\smallskip} 
\multicolumn{9}{c}{Supergiants}\\
\noalign{\smallskip}
\tableline
HD 123008            & 14:07:30.650 & $-$64:28:08.82  & ON9.5 Iab     &8.837 
& +0.371  & $-$6.5  & 30500  & $-$2.9\\ 
HD 191781            & 20:09:50.581 & +45:24:10.44  & ON9.7 Iab       &9.533 
& +0.636  & $-$6.5  & 30050 & $-$2.8\\ 
HD 122879            & 14:06:25.157 & $-$59:42:57.25  & B0 Ia         &6.410 
& +0.124  & $-$7.0  & 29600 & $-$2.8\\ 
HD 148546            & 16:30:23.312 & $-$37:58:21.15  & O9 Iab        &7.711 
& +0.291  & $-$6.5  & 31400  & $-$3.0\\ 
HD 149038            & 16:34:05.023 & $-$44:02:43.14  & O9.7 Iab      &4.910 
& +0.078  & $-$6.5  & 30050 & $-$2.8\\  
HD 154368            & 17:06:28.371 & $-$35:27:03.76  & O9.5 Iab      &6.133 
& +0.509  & $-$6.5  & 30500  & $-$2.9\\ 
HD 188209            & 19:51:59.068 & +47:01:38.44  & O9.5 Iab        &5.625 
& $-$0.064  & $-$6.5  & 30500  & $-$2.9\\ 
HD 195592            & 20:30:34.970 & +44:18:54.87  & O9.7 Ia         &7.080 
& +0.870  & $-$7.0  & 30050 & $-$2.8\\ 
HD 202124            & 21:12:28.389 & +44:31:54.14  & O9 Iab          &7.813 
& +0.230  & $-$6.5  & 31400  & $-$3.0\\ 
HD 104565            & 12:02:27.795 & $-$58:14:34.36  & OC9.7 Iab     &9.265 
& +0.363  & $-$6.5  & 30050 & $-$2.8\\ 
HD 152249            & 16:54:11.641 & $-$41:50:57.27  & OC9 Iab       &6.463 
& +0.202  & $-$6.5  & 31400  & $-$3.0\\ 
HD 152424            & 16:55:03.331 & $-$42:05:27.00  & OC9.5/9.7 Ia  &6.311 
& +0.400  & $-$7.0  & 30500  & $-$2.9\\  
HD 154811            & 17:09:53.086 & $-$47:01:53.19  & OC9.7 Iab     &6.921 
& +0.394  & $-$6.5  & 30050 & $-$2.8\\ 
HD 194280            & 20:23:26.313 & +38:56:21.01  & BC0 Iab         &8.390 
& +0.760  & $-$6.5  & 29600 & $-$2.8\\ 
\tableline
\noalign{\smallskip}                                                            
                       
\multicolumn{9}{c}{Giants}\\
\noalign{\smallskip}
\tableline                                                                      
                         
HD 13268             & 02:11:29.700 & +56:09:31.70  & ON8.5 IIIn      &8.182 
& +0.128  & $-$5.6  & 32900  & $-$3.1\\ 
HD 89137             & 10:15:40.086 & $-$51:15:24.08  & ON9.7 II--III(n) 
& 7.974 & $-$0.045  & $-$5.5  & 30250 & $-$2.8\\ 
HD 91651             & 10:33:30.301 & $-$60:07:40.04  & ON9.5 IIIn    &8.849 
& $-$0.008  & $-$5.3  & 30800  & $-$2.9\\ 
HD 102415            & 11:46:54.404 & $-$61:27:46.99  & ON9 IIInn     &9.146 
& +0.120  & $-$5.6  & 31800  & $-$3.0\\ 
HD 117490            & 13:32:08.600 & $-$60:48:55.47  & ON9.5 IIInn   &8.902 
& +0.033  & $-$5.3  & 30800  & $-$2.9\\ 
HD 150574            & 16:44:07.209 & $-$46:08:29.85  & ON9 III(n)    &8.497 
& +0.232  & $-$5.6  & 31800  & $-$3.0\\ 
HD 191423            & 20:08:07.113 & +42:36:21.98  & ON9 II--IIInn   &8.030 
& +0.160  & $-$5.75 & 31700  & $-$3.0\\
\end{tabular}
\end{table}

\addtocounter{table}{-1}
\begin{table}
\caption{Continued} 
\vspace{0.2cm}
\begin{tabular}{llrlcrlll}
\tableline
\tableline
\noalign{\smallskip}
\multicolumn{1}{c}{Name}&R.A.(2000.0)&Decl(2000.0)&Spectral
Type&$V$&$B-V$&\multicolumn{1}{c}{$M_V$}&\multicolumn{1}{c}{$T_{\rm
eff}$}&\multicolumn{1}{c}{BC}\\
\noalign{\smallskip}
\tableline
\noalign{\smallskip} 
HD 10125             & 01:40:52.762 & +64:10:23.13  & O9.7 II        &8.220 
& +0.310  & $-$5.9  & 30200 & $-$2.8\\ 
HD 15137             & 02:27:59.811 & +52:32:57.60  & O9.5 II--IIIn  &7.870 
& +0.030  & $-$5.6  & 30650  & $-$2.9\\ 
HD 24431             & 03:55:38.420 & +52:38:28.75  & O9 III         &6.732 
& +0.371  & $-$5.6  & 31800  & $-$3.0\\ 
HD 93521             & 10:48:23.511 & +37:34:13.09  & O9.5 IIInn     &7.040
& $-$0.280  & $-$5.3  & 30800  & $-$2.9\\
HD 96264             & 11:04:55.501 & $-$61:03:05.79  & O9.5 III     &7.606 
& $-$0.064  & $-$5.3  & 30800  & $-$2.9\\ 
HD 114737            & 13:13:45.528 & $-$63:35:11.75  & O8.5 III     &7.995 
& +0.172  & $-$5.6  & 32900  & $-$3.1\\ 
HD 154643            & 17:08:13.983 & $-$35:00:15.68  & O9.7 III     &7.165 
& +0.278  & $-$5.1  & 30300 & $-$2.8\\ 
HD 189957            & 20:01:00.005 & +42:00:30.83  & O9.7 III       &7.806 
& +0.013  & $-$5.1  & 30300 & $-$2.8\\ 
HD 207198            & 21:44:53.278 & +62:27:38.05  & O9 II          &5.943 
& +0.311  & $-$5.9  & 31600  & $-$3.0\\ 
\tableline
\noalign{\smallskip}
\multicolumn{9}{c}{Dwarfs}\\
\noalign{\smallskip}
\tableline 
HD 12323             & 02:02:30.126 & +55:37:26.38  & ON9.5 V        &8.900 
& $-$0.092  & $-$4.1  & 31900  & $-$3.0\\  
HD 14633             & 02:22:54.293 & +41:28:47.72  & ON8.5 V        &7.458 
& $-$0.212  & $-$4.4  & 33900  & $-$3.2\\ 
HD 48279             & 06:42:40.548 & +01:42:58.23  & ON8.5 V        &7.910 
& +0.136  & $-$4.4  & 33900  & $-$3.2\\ 
HD 201345            & 21:07:55.416 & +33:23:49.25  & ON9.5 IV       &7.660 
& $-$0.130  & $-$4.7  & 31350  & $-$3.0\\ 
HD 46149             & 06:31:52.533 & +05:01:59.19  & O8.5 V         &7.601 
& +0.171  & $-$4.4  & 33900  & $-$3.2\\ 
HD 46202             & 06:32:10.471 & +04:57:59.79  & O9.5 V         &8.182 
& +0.177  & $-$4.1  & 31900  & $-$3.0\\ 
HD 93028             & 10:43:15.340 & $-$60:12:04.21  & O9 IV        &8.361 
& $-$0.071  & $-$5.0  & 32350  & $-$3.0\\  
HD 93027             & 10:43:17.954 & $-$60:08:03.29  & O9.5 IV      &8.720 
& $-$0.020  & $-$4.7  & 31350  & $-$3.0\\ 
HD 149757            & 16:37:09.530 & $-$10:34:01.75  & O9.5 IVnn    &2.565
& +0.019  & $-$4.7  & 31350  & $-$3.0\\
HD 214680            & 22:39:15.679 & +39:03:01.01  & O9 V           &4.879 
& $-$0.201  & $-$4.3  & 32900  & $-$3.1\\  
\tableline
\end{tabular}
\end{table}

\begin{table}
\caption{Velocities, Galactic Latitudes, and Nitrogen Abundances for ON
Stars}
\medskip
\begin{tabular}{llclcrrr}
\tableline
\tableline
\noalign{\smallskip}
\multicolumn{1}{c}{Name} &\multicolumn{1}{c}{Spectral Type\tablenotemark{a}} 
&$v\sin i$\/\tablenotemark{b} &\multicolumn{1}{c}{$v_r$} 
&\multicolumn{1}{c}{Source\tablenotemark{c}}
&\multicolumn{1}{c}{$b$} &\multicolumn{1}{c}{N\tablenotemark{d}} 
&\multicolumn{1}{c}{Source\tablenotemark{e}}\\
&&\multicolumn{1}{c}{[km s$^{-1}$]} &\multicolumn{1}{c}{[km s$^{-1}$]} 
&&\multicolumn{1}{c}{[deg]}\\
\noalign{\smallskip}
\tableline
\noalign{\smallskip}
\multicolumn{7}{c}{Supergiants}\\
\noalign{\smallskip}
\tableline
HD 123008 &ON9.5 Iab &98 &$-$21, $-$36 C &1, 2 &$-$2.8 &$\geq$1.8 &1\\
HD 191781 &ON9.7 Iab &\llap{$\sim$1}00 &$-$13, $-$10 V? &3, 4 
&+6.6 &...\phn &...\\
\tableline
\noalign{\smallskip}
\multicolumn{7}{c}{Giants}\\
\noalign{\smallskip}
\tableline
HD 13268 &ON8.5 IIIn  &\llap{3}09 &$-$127, $-$123\phn C &5, 6 
&$-$5.0 &...\phn &...\\
HD 89137 &ON9.7 II--III(n) &\llap{2}02 &+3, $-$25 V? &1, 2 &+4.4 
&25\hphantom{.0} &2\\
HD 91651 &ON9.5 IIIn &\llap{2}92 &$-$41 SB2 &1, 2 &$-$1.7 &...\phn &...\\
HD 102415 &ON9 IIInn &\llap{$\sim$4}00 &$-$10 V? &7 &+0.4 &...\phn &...\\
HD 117490 &ON9.5 IIInn &\llap{$\sim$4}00 &+3 SB2? &7 &+1.7 &...\phn &...\\
HD 150574 &ON9 III(n)  &\llap{$\sim$2}00 &$-$56 SB2 &2 &$-$0.2 &...\phn &...\\
HD 191423 &ON9 II--IIInn &\llap{4}35 &$-$38, $-$90 V? &3, 8 &+5.4 &3.1 &3\\
\tableline
\noalign{\smallskip}
\multicolumn{7}{c}{Dwarfs}\\
\noalign{\smallskip}
\tableline
HD 12323 &ON9.5 V &\llap{1}31 &$-$42 SB &4, 6 &$-$5.9 &$\geq$2.8 &1\\
HD 14633 &ON8.5 V &\llap{1}34 &$-$38 SB1 &9 &$-$18.2 &$\geq$2.8, 13 &1, 3\\
HD 48279 &ON8.5 V &\llap{1}28 &+29 C &2, 4 &$-$1.2 &$\geq$2.0, 16 &1, 2\\
HD 201345 &ON9.5 IV &91 &+20 SB2? &4 &$-$9.5 &33/6.5\rlap{,\tablenotemark{f}} 
&4\rlap{,}\\
&&&&&&$\geq$2.0, 8 &1, 5
\end{tabular}
%
\tablenotetext{a}{HD~123008: revised from ON9.7 to ON9.5 in southern
GOSSS in prep., but foreseen by Walborn \& Fitzpatrick 1990.
HD~13268: ON discovered by Mathys 1989.  HD~89137: pec.\ in Walborn
1976, 2003 and N str.\ in Garrison et~al.\ 1977; ON confirmed by GOSSS.
HD~91651: N str.\ in Garrison et~al.\ 1977; ON confirmed by GOSSS.
HD~102415, 117490: ON discovered by southern GOSSS.  HD~191423: ON
discovered by Howarth \& Smith 2001.  HD~12323, 201345: revised from ON9
to ON9.5 and latter from V to IV in GOSSS.  HD~14633, 48279: revised from
O(N)8 to O(N)8.5 in GOSSS and latter from N str.\ (Walborn 1976) to ON in
Sana, Sim\'on-D\'{\i}az, Walborn et~al.\ in prep.}       
\tablenotetext{b}{Projected rotational velocities from Howarth et~al.\ 1997 
or Howarth \& Smith 2001 as available; otherwise inferred approximately from 
the classification line-broadening parameter.  Note the perfect agreement with 
measured values.  The actual rotational velocities may be lower if additional 
line-broadening mechanisms are significant (Sim\'on-D\'{\i}az \& Herrero 2007).}
\tablenotetext{c}{1: Feast et~al.\ 1963. 2: Levato et~al.\ 1988.  
3: Petrie \& Pearce 1961. 4: Bolton \& Rogers 1978. 5: Abt et~al.\ 1972. 
6: Kendall et~al.\ 1995. 7: R.~Gamen et~al.\ 2011, in prep.\  
8: Howarth et~al.\ 1997. 9: Boyajian et~al.\ 2005.  
$\gamma$ for SB.  C:~constant, V:~variable.}
\tablenotetext{d}{Nitrogen abundance factors relative to 10~Lac, except 
Source~1 to ``normal stars.''}
\tablenotetext{e}{1: Wollaert et~al.\ 1988. 2: Sch\"onberner et~al.\ 1988; 
5: 1984. 3: Villamariz et~al.\ 2002. 4: Lester 1973.}
\tablenotetext{f}{First Source 4 value is an average of nine lines ranging over 
two orders of magnitude; second value omits four anomalous, excessive lines.}
\end{table}

\section{Discussion}
\subsection{Binarity}

HD~12323, HD~14633, and HD~201345 have been reported as definite or
probable spectroscopic binaries by Lester (1973), Bolton \& Rogers
(1978), and/or Boyajian et~al.\ (2005).  HD~48279 was found to have
constant radial velocity by Bolton \& Rogers (1978), Levato et~al.\  
(1988), and Mahy et~al.\ (2009).  Indeed, binary mass transfer would 
appear the most likely origin of dwarf or subgiant ON stars.  
It could also be a possible explanation of the ONn class in principle 
(Langer et~al.\ 2008).  However, their binary status is currently unclear.
Levato et~al.\ reported HD~89137 as probably variable, and HD~91651 and
HD~150574 as probably SB2, the last based on the report of double
lines by Garrison, Hiltner, \& Schild (1977).  HD~13268 was found to have
a constant radial velocity by Abt, Levy, \& Gandet (1972), but see below.
A single epoch with double lines was reported for HD~191423 by Petrie \&
Pearce (1961), which definitely requires observational followup.
The new ONn stars HD~102415 and HD~117490 are likely velocity variables.
These results are summarized in Table~2, which shows that 10 of the 13 ON
stars listed are definite or possible spectroscopic binaries.

Several high-resolution radial-velocity programs on the O stars are in 
progress, and the ONn class is an outstanding candidate for further data.
The preliminary results in Table~2 for the new ONn stars are from the extensive 
OWN Survey led by R.C.G.\ and R.H.B.\ (Barb\'a et~al.\ 2010; Arias et~al.\ 2010). 
Of course, given the high incidence of multiplicity among massive stars, 
the mere fact of binarity does not prove that mass transfer has taken place.  
A further complication is that some binaries may merge, producing a single, 
rapidly rotating object.

\subsection{Space Distribution and Motions}

None of the ON or ONn stars listed in Tables~1 and 2 is a definite cluster or
association member.  On the contrary, as given in Table~2, many of them 
have unusually high galactic latitudes for OB stars and several have high 
radial velocities, as previously discussed by Walborn (1970), Bolton \& Rogers 
(1978), Kendall et~al.\ (1995), Howarth et~al.\ (1997), and Boyajian et~al.\  
(2005).  
  
The extreme radial velocity of HD~13268 was found by Abt et~al.\ 
(1972), its ON nature by Mathys (1989), and both were confirmed and further 
discussed by Kendall et~al.\ (1995).  Abt et~al.\ and Kendall et~al.\ reasonably 
suggested that HD~13268 is a runaway star, as has been proposed for other 
ON stars in the references cited in the previous paragraph.  Indeed,
its position and proper motions are quite similar to those of HD~14633
and HD~15137, for both of which an origin in NGC~654 was suggested by 
Boyajian et~al.\  HD~12323 also bears comparison in these properties.  
Clearly this is an important consideration with regard to the origin of 
these objects, as further discussed below.

\subsection{Abundances}

Unfortunately, quantitative CNO abundance determinations for the ON stars 
are sparse and dated.  Relative nitrogen abundances as available from the
literature are given in Table~2, for reference in the discussion below.  
However, only that for HD~191423 (Villamariz et~al.\ 2002) has been derived 
with current techniques.  Clearly, state-of-the-art abundance studies of all 
these stars are essential for substantive progress in their interpretation.  
The helium abundances should also be investigated systematically.  Again,
Table~1 here provides an optimum sample for comparative analysis of the
relevant categories.

\subsection{Evolution}

\subsubsection{Masses, Ages, and N Enhancements}

An HR Diagram for the ON stars is presented in Figure~2.  It displays
evolutionary tracks with solar composition for nonrotating (Schaller et~al.\  
1992) and 300~km~s$^{-1}$ initially rotating (Meynet \& Maeder 2000) models
of 20, 25, 40, and 60~$M_{\sun}$.  The rotating tracks are labeled with 
nitrogen abundances relative to the ZAMS.  Nonrotating isochrones for   
3.2 and 5.6~Myr are also shown.  Detailed comparisons of spectral-type
calibrations on these tracks were discussed by Walborn \& Lennon (2003).
The differences between the rotating and nonrotating cases are relatively
small in the domain of the ONn stars, as also seen here; but significant
mixing of N is not predicted by the nonrotating models at these stages.

\begin{figure}
\epsscale{1.0}
\plotone{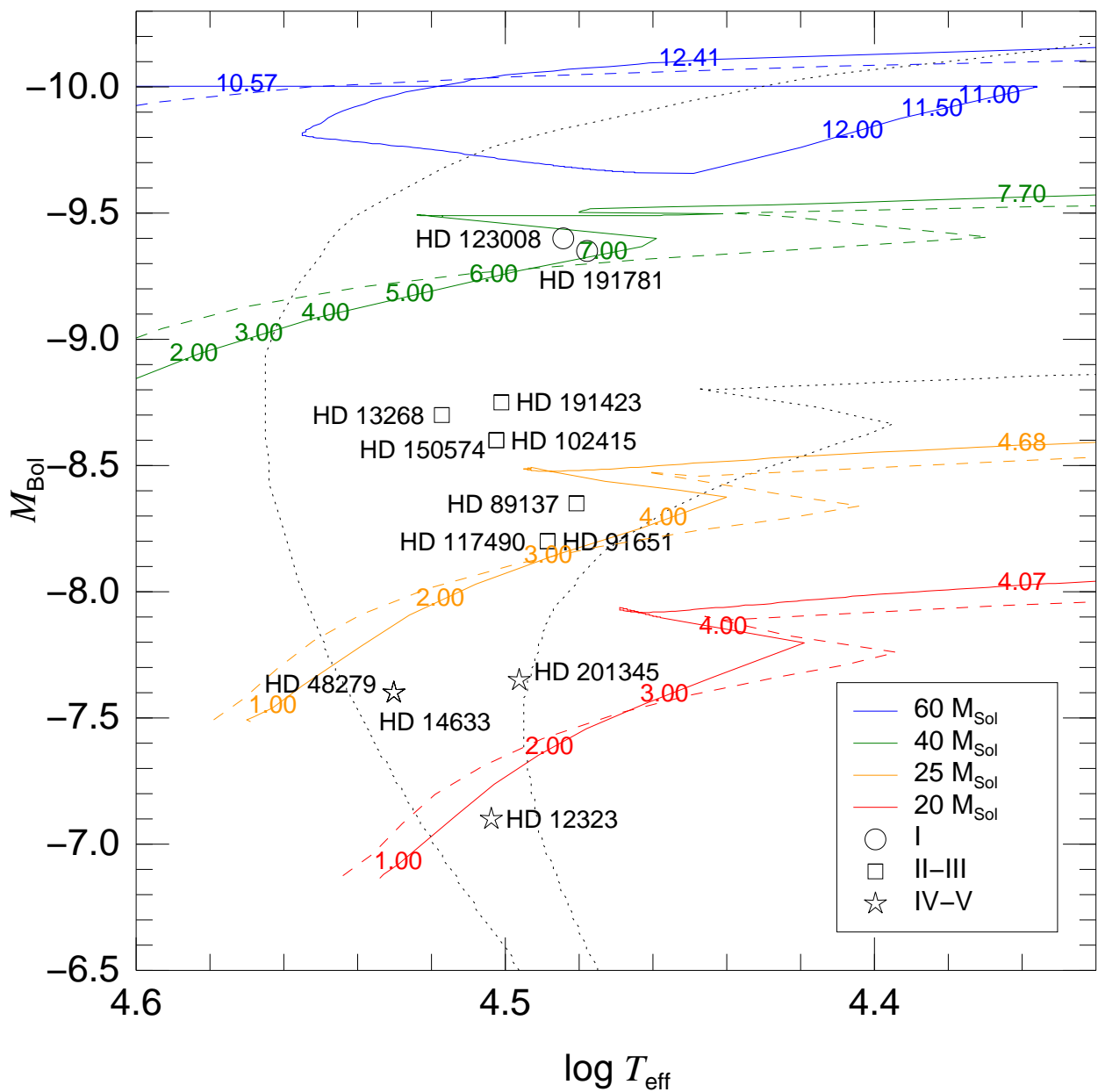}
\caption{
An HR Diagram for the ON stars.  The ONn stars are the open squares
(luminosity classes III--II).  Note that the locations of stars with
identical two-dimensional classifications are degenerate and shown as
single points.  The dashed tracks are for nonrotating and the solid 
tracks for rotating models; the latter are labeled with surface N
abundances relative to the ZAMS value.  The dotted lines are nonrotating 
isochrones for 3.2 and 5.6~Myr.}
\end{figure}

The stars have been plotted in Figure~2 based on the spectral-type
calibrations in Table~1.  By that procedure, any star with a given
two-dimensional classification appears at the same point in the diagram,
regardless of N or rotational differences.  Thus, it is not possible to
investigate any systematic effects of those parameters in this way.
Also, should the ON stars be products of binary mass transfer, the
calibrations for normal stars may not be accurate for them.
Reliable individual quantitative analyses and/or distances, which do not 
exist for most of these stars at present but will in the future, are
required for more definitive results and comparisons.  Although the normal 
and OC comparison stars are hence not discussed further here, they are 
retained in the table as optimal comparisons for the ON stars in future work.

According to the HRD, the ONn stars have masses in the range of 
25--30~$M_{\sun}$ and ages of 4--5~Myr.  The ON dwarfs and subgiant are 
in the 20--25~$M_{\sun}$ range, while the two ON supergiants lie on the 
40~$M_{\sun}$ track.  Thus, these subcategories are not evolutionarily 
related; rather they all have similar spectral types and effective 
temperatures at least partly as a consequence of the selective 
detectability of the CNO anomalies at those types.  Quantitative work 
will be required to identify their hotter progenitors, which should have 
main-sequence spectral types in the O7--O8 range.  HD~110360 has a very 
clear ON7~Vz type, as originally discovered by Mathys (1989) and confirmed 
by further GOSSS work in progress, in the course of which the ``z,'' possibly 
related to extreme youth (Walborn 2009b) has been added; however, it has very 
sharp lines and thus likely a low rotational velocity.  Some descendant 
analogues of the ONn stars may be found among the BN class (Walborn 1976). 

Then, one would like to compare the observed nitrogen enhancements (Table~2) 
with those predicted by the models.  Unfortunately, only the observational 
result for HD~191423 is likely to be sufficiently reliable for a meaningful 
comparison.  Its observed value is in excellent agreement with the 
prediction for its HRD location, supporting an internal origin of its 
current N abundance, as was already concluded by Villamariz et~al.\ (2002).
From the strong spectroscopic similarities, it is reasonable to expect
that similar results will obtain for the other ONn stars.  On the other
hand, as noted above, a possible binary nature of HD~191423 requires
further investigation, both to confirm it or otherwise, and to elucidate
any relationship to the N enhancement if confirmed; and similarly for the
other ONn objects.

\subsubsection{Origins}

Several alternative interpretations of the ONn stars may be considered in
principle and compared with the available observational and theoretical
constraints.

\noindent (i) Stars with rapid initial rotations are expected to undergo
more substantial mixing of processed material to their surfaces (Maeder
\& Meynet 2000; Heger \& Langer 2000).  In extreme cases they may evolve
homogeneously back toward the ZAMS (Langer 1992; Meynet \& Maeder 2000).  
However, the question of braking by the stellar winds then arises; are 
the current rotational velocities of the ONn stars compatible with even 
higher initial values and the expected wind braking?  Unfortunately, the 
mass-loss rates of late-O dwarfs are currently highly uncertain, with 
models predicting larger values than implied by observations (Marcolino 
et~al.\ 2009), so a definitive answer to this question is not at hand.

\noindent (ii) Core contraction at the TAMS stage may increase the
rotational velocities, and the ONn stars are near or at that stage.  
However, so are the far more numerous normal stars of the same spectral 
types, which argues against that mechanism.

\noindent (iii) Mass transfer in a binary system can produce both
rotational acceleration and processed material on the surface of the
secondary, whether from the primary or from induced mixing in the
secondary (Langer et~al.\ 2008).  In extreme cases, a merger may be the
end result.  In either event, homogeneous evolution may then ensue.
If the original primary has already undergone a SN explosion, the secondary 
or even the remnant binary may acquire a high space velocity, which is a 
property of at least some ON stars as discussed above.  Blaauw (1993) 
summarized evidence for helium enhancements in some other runaway stars in 
this context.  High helium abundances in several ON stars have been derived 
by Sch\"onberner et~al.\ (1988), Herrero et~al.\ (1992), Smith \& Howarth
(1994), Kendall et~al.\ (1995), and Howarth \& Smith (2001).
In this case, the discussion of single-star evolution becomes irrelevant, 
which could also be compatible with the rarity of the ON stars. 

It is difficult to choose among these alternative interpretations at
the present time, except for the likely elimination of (ii).  However,
observational and theoretical developments currently underway may be
expected to improve this situation in the relatively near future.

\newpage

\section{Summary}

The ONn category of nitrogen-rich, rapidly rotating late-O giants has
been expanded and enhanced with improved data from the GOSSS.  In
particular, two southern-hemisphere ``clones'' (HD~102415 and HD~117490) 
of the most rapidly rotating ON star previously known (HD~191423) have been 
added to the category.  The properties of the class as currently available 
have been discussed, including rotation, binarity, space distribution and 
motions, and nitrogen abundances.  A comparison with rotating stellar models 
has been made.  However, the observational data are fragmentary and in some 
cases inadequate, with the result that alternative interpretations of the 
ONn class in terms of single-star or binary evolution cannot be definitively
distinguished, although the latter may be somewhat favored.  This sample 
of stars, including an extensive list of normal and OC comparison objects, 
provides an outstanding project for further observations and analysis, 
some of which are already in progress. 

\acknowledgments
N.R.W.\ thanks the Sociedad Espa\~nola de Astronom\'{\i}a for generous
international travel and subsistence support, and the Instituto de
Astrof\'{\i}sica de Andaluc\'{\i}a and its staff for kind hospitality 
and subsistence (through Spanish Government grants AYA2007-64052 and
AYA2010-17631), during 2010 Sept.--Oct.  Ancillary support, including 
publication charges, was provided by NASA through grants GO-10898.01 and
GO-12179.01 from STScI, which is operated by AURA, Inc., under NASA contract 
NAS5-26555.  J.M.A, A.S., and E.J.A.\ acknowledge support by the Spanish 
Government Ministerio de Ciencia e Innovaci\'on grants AYA2007-64052 
and AYA2010-17631, and by the Junta de Andaluc\'{\i}a grant P08-TIC-4075; 
J.M.A.\ was also supported by the Ram\'on y Cajal Fellowship program, 
cofinanced by the European Regional Development Fund (FEDER).  
R.H.B.\ acknowledges partial support from Universidad de La Serena Project 
DIULS CD08102.  We are grateful for generous allocations of observing time 
at the Observatorio de Sierra Nevada and the Las Campanas Observatory.
Thanks to the anonymous referee for some interesting suggestions that led
to additional relevant information.

\end{document}